\documentstyle[12pt]{article}
\begin{document}
\begin{center}
\vskip 3.5\baselineskip
{\bf \Large Spin Structure of Nucleon and Equivalence Principle}
\vskip 2.5\baselineskip
O.V. Teryaev$^{1}$
\vskip \baselineskip
1. Bogoliubov Laboratory of Theoretical Physics, JINR, \\
   141980 Dubna, Moscow region, Russia\
\vskip 3\baselineskip
{\bf Abstract} \\[0.5\baselineskip]
\parbox{0.9\textwidth}{
The partition of nucleon spin between total angular momenta
of quarks and gluons is described by the energy momentum tensor
formfactors manifested also in the nucleon scattering by weak classical
gravitational field. Natural generalization of equivalence principle
is resulting in the identically zero "Anomalous gravitomagnetic moment"
being the straightforward analog of its electromagnetic
counterpart. This, in turn, means the equal partition of
momentum and total angular momentum, anticipated earlier.}

\vskip 1.5\baselineskip
\end{center}


\noindent
{\bf 1 Introduction.}

Spin structure of the nucleon is known to be one of the
most intriguing problems of non-perturbative QCD \cite{AEL}.
Its solution happened to deal with the fine details
of the theory. In particular, the gluon spin momentum is
intimately related to the axial anomaly, entirely
responsible for the first moment of the relevant distribution
\cite{anom}  and manifested also as an
$x$-dependent term \cite{MT}.

The orbital angular momenta are the necessary counterparts of the spin one,
required already by the leading QCD evolution \cite{Ratcl,Ji,HS}.
The current mainstream of their studies is following the suggestion
of X. Ji \cite{Ji} to use the relation of {\it total}
angular momenta to the particular matrix elements of Belinfante
energy momentum tensors.

\begin{eqnarray}
      \langle p'| T_{q,g}^{\mu\nu} |p\rangle
       &=& \bar u(p') \Big[A_{q,g}(\Delta^2)
       \gamma^{(\mu} p^{\nu)} +
   B_{q,g}(\Delta^2) P^{(\mu} i\sigma^{\nu)\alpha}\Delta_\alpha/2M ] u(p)
\label{def}
\end{eqnarray}
where $P^\mu=(p^\mu+{p^\mu}')/2$, $\Delta^\mu = {p^\mu}'-p^\mu$,
and $u(p)$ is the nucleon spinor.
We dropped here the irrelevant terms of higher order in $\Delta$,
as well as containing $g^{\mu \nu}$, which will be discussed later.
The parton momenta and total angular momenta are just
\begin{eqnarray}
      P_{q, g} = A_{q,g}(0), \nonumber \\
      J_{q, g} = {1\over 2} \left[A_{q,g}(0) + B_{q,g}(0)\right] \ .
\end{eqnarray}
Taking into account the conservation of momentum
and angular momentum
\begin{eqnarray}
A_q(0)+A_g(0)= 1 \label{m} \\
A_q(0)+B_q(0)+A_g(0)+B_g(0)=1 \label{am}
\end{eqnarray}

one can see, that the
difference between partition of momentum and orbital angular momentum
is entirely coming from "anomalous" formfactors
($B_q(0)=-B_g(0)$).

The smallness of such a contributions comes from
the models \cite{Man}, QCD sum rules calculations \cite{BJ} and
chiral soliton models \cite{Boh}. The careful analysis
and rederivation \cite{OT98}
of the leading QCD evolution \cite{HS} was used to make a statement
about the {\it identical} zero of the anomalous formfactor $B$.
Moreover, just this general property was suggested in \cite{OT98}
as a reason for the smallness of the singlet anomalous magnetic moment,
resulting from the approximate cancellation between proton ($+1.79$)
and neutron ($-1.91$) values.

To prove this picture one should either study the higher orders
and (especially) nonperturbative QCD contributions, or to look
for a more general reason. In the present paper the latter
approach is suggested.
Namely, making use of the fact that
the matrix element (\ref{def}) is describing the interaction
of nucleon with the classical external gravitational field
one arrive to the interpretation of $B$ as an
"Anomalous Gravitomagnetic Moment" (AGM), being the straightforward
analog of its electromagnetic counterpart.
The natural extension of the famous Einstein equivalence principle
is resulting in the zero AGM. As a byproduct, one
can see, that the helicity of any Dirac particle (say, massive
neutrino), is not flipped by the rotation of astronomical objects.

\vskip\baselineskip
\noindent
{\bf 2 Nucleon in the external gravitational field.}

Let us start with the more common case of the interaction
with electromagnetic field, described by the matrix element
of electromagnetic current,
\begin{eqnarray}
  M=\langle P'| J_{q}^{\mu} |P\rangle A_\mu.
\label{A}
\end{eqnarray}
This matrix element at zero momentum transfer is fixed by the
fact, that the interaction is due to the {\it local} U(1) symmetry,
whose {\it global} counterpart is producing the conserved charge
(and of course is depending on the normalization of eigenvectors
$\langle P|P'\rangle=(2\pi)^3 2E \delta(\vec P-\vec P')$).

\begin{eqnarray}
\langle P| J_{q}^{\mu} |P\rangle=2e_q P^\mu,
\label{e}
\end{eqnarray}
so that in the rest frame the interaction is completely defined by
the scalar potential:
\begin{eqnarray}
\label{0e}
  M_0=\langle P| J_{q}^{\mu} |P\rangle A_\mu = 2e_q M \phi
\label{A}
\end{eqnarray}
At the same time, the interaction with the weak classical gravitational
field is:
\begin{eqnarray}
M=\frac{1}{2}\sum_{q,G} \langle P'| T_{q,G}^{\mu} |P\rangle h_{\mu \nu},
\label{h}
\end{eqnarray}
where $h$ is a deviation of metric tensor from its Minkowski value.
The relative
factor 1/2, which will play a crucial role, is coming from the fact,
that the variation of the action with respect to the metric
is producing an energy-momentum tensor with the coefficient 1/2,
while the variation with respect to classical source $A^\mu$,
is producing the current without such a coefficient. It is this
coefficient, that guarantee the correct value for the
Newtonian limit, fixed by the {\it global} translational invariance
\begin{eqnarray}
\sum_{q,G} \langle P| T_{i}^{\mu \nu} |P\rangle=2 P^\mu P^\nu,
\label{e}
\end{eqnarray}
which, together with the approximation for $h$
(with factor of 2
having the geometrical origin) \cite{LL2}
\begin{eqnarray}
\label{h}
h_{00}=2\phi (x)
\end{eqnarray}
is resulting in the rest frame expression:
\begin{eqnarray}
\label{0g}
M_0=\sum_{q,G} \langle P| T_{i}^{\mu \nu} |P\rangle h_{\mu \nu}=
2 M \cdot M \phi,
\label{A}
\end{eqnarray}
where we used the same notation for gravitational and scalar
electromagnetic potentials, and identified normalization factor
$2M$ in order to make the similarity between (\ref{0e}) and (\ref{0g})
obvious. One can see that the interaction with gravitational
field is described by the charge , equal to the particle mass,
which is just the equivalence principle.
It is appearing here as low energy theorem, rather than postulate.
The similarity with
electromagnetic case allows to clarify the origin of such a theorem,
suggesting, that the interaction with gravity is due to the {\it local}
counterpart of {\it global} symmetry, although it may be proved
starting just from the Lorentz invariance of the soft graviton
approximation \cite{wein64}.

The situation with the terms linear in $\Delta$ is different
for electromagnetism and gravity. While such a term is
defined by the specific dynamics in the electromagnetic case,
producing the anomalous magnetic moment, the similar contribution
in the gravitational case is entirely fixed by the angular momentum
conservation (\ref{am}),
which
was known in the context of gravity for more than 20 years \cite{BD,HD}.
\footnote{The reason is that the structure of
Poincare group
is more reach than
that of $U(1)$ group.}.
It
means,
in terms of the gravitational interaction, that
{\bf Anomalous Gravitomagnetic Moment (AGM) of any particle
is identically equal to
zero}.

Let us clarify this statement, which
is not restricted to the nucleon or spin-1/2 Dirac
particle. The presence of Dirac spinors in the
parametrization (\ref{def}) is actually not crucial.
To show that, it is convenient to use the equation of
motion in order to attribute all the $\Delta$-dependence to the
{\it anomalous} formfactor
$P^\mu \bar u \sigma^{\nu \alpha} u \Delta_\alpha$.
As soon as the linear $\Delta$-dependence is already
extracted, the spinors can be taken at the same momentum,
which is convenient to choose as an average one $P$,
and calculation of the matrix element is reduced to the
trace of density matrix:
\begin{eqnarray}
\bar u(P) \sigma^{\nu \alpha} \Delta_\alpha u(p)=
Tr \rho(P) i\sigma^{\nu \alpha}\Delta_\alpha =  \nonumber \\
Tr \frac{1}{2}(\hat P+M)(1+\hat S \gamma_5)
i\sigma^{\nu \alpha} \Delta_\alpha
=2 i \epsilon^{\rho \sigma \nu \alpha}
P^\rho S^\sigma \Delta^\alpha
\label{gen}
\end{eqnarray}

The constraint (\ref{am}), by considering
the matrix element of the projection of
Pauli-Lubanski operator, may be now easily generalized
to the particle of {\it any} spin, so that, for the {\it total}
conserved energy momentum tensor of all the constituents is
(c.f. \cite{JM}):
\begin{eqnarray}
      \langle P'| \sum T^{\mu\nu} |P\rangle
       &=& 2 P^{\mu} P^{\nu} + i P^{(\mu}\epsilon^{ \nu)  \sigma \rho \alpha}
P^\rho S^\sigma \Delta^\alpha/M.
\label{genc}
\end{eqnarray}
Like in the spin 1/2 case, $S$ is the average spin in any of the
states $|P>,|P'>$ (the difference between which is inessential,
as soon as linear terms in $\Delta$
are considered, and we postpone the discussion
of the spin-flip case.)

As soon as the formfactors in spin-1/2 case differ from the ones for
the matrix element of vector current $J^\mu$ by the common factor
$P^\nu$, one may define
{\it gyrogravitomagnetic ratio} in the same way as common
gyromagnetic ratio, and it should have Dirac value $g=2$
for particle of any spin $J$:
\begin{eqnarray}
\mu_G=J
\label{genc}
\end{eqnarray}
which coincide with the standard Dirac magnetic moment,
up to the interchange $e \leftrightarrow M$,
making the Bohr magneton equal to $1/2$.

However, the situation changes if one define the gyrogravitomagnetic moment
as a response to the external gravitomagnetic field.
The $\epsilon$ tensor in the coordinate space produce the curl,
and the gravitomagnetic field, acting on the particle spin, is equal to
\begin{eqnarray}
\vec H_J = \frac{1}{2} rot \vec g; \ \vec g_i \equiv g_{0i},
\label{hg}
\end{eqnarray}
where factor $1/2$ is just the mentioned normalization factor in
(\ref{h}). The relevant off-diagonal components of the metric
tensor may be generated by the rotation of
massive gravity source \cite{LL2}.

There is also another effect, induced by this field:
the straightforward analog
of Lorentz force \cite{LL2}, produced by the first (spin-independent) term
in (\ref{genc}). In that case the gravitomagnetic filed,
for the low velocity of the particle (such a restriction is actually
inessential, as we can always perform the Lorentz boost, making
the particle velocity small enough) is:

\begin{eqnarray}
\vec H_L = rot \vec g = 2 \vec H_G,
\label{hg}
\end{eqnarray}

Consider now the motion of the particle in the gravitomagnetic field.
The effect of Lorentz force is reduced, due to the Larmor theorem,
(which is also valid for small velocity)
to the rotation with the Larmor frequency
\begin{eqnarray}
\omega_L= \frac{H_L}{2}.
\label{hg}
\end{eqnarray}
This is also the frequency of the {\it macroscopic} gyroscope dragging.
At the same time, the {\it microscopic} particle dragging frequency is
\begin{eqnarray}
\omega_J= \frac{\mu_G}{J}H_J=\frac{H_L}{2}=\omega_L.
\label{hg}
\end{eqnarray}
The common frequency for microscopic and macroscopic gyroscope
is just the Larmor frequency, so that the gravitomagnetic field
is equivalent to the frame rotation. This should be considered as
a Post-Newtonian manifestation of the equivalence principle.

Let us make here a brief comparison with the literature.
The low energy theorem discussed here
is the necessary ingredient for validity
of gravitational Larmor theorem \cite{Mash}, which otherwise
require an arbitrary assumption about the "classical"
gyrogravitomagnetic
ratio, say, for electron \cite{Mash2}.
At the same time, the equality
of the "classical" and "quantum" frequencies was
found long ago \cite{HD} by comparison of the quantum spin-orbit interaction
with the classical calculated earlier \cite{Sch}.
Our approach clarify the origin of this equality, as a cancellation
of "geometrical" factor $1/2$ in (\ref{h}) and "quantum"
value $2$ of gyrogravitomagnetic ratio. Note that for free
particle the latter coincides with the usual gyromagnetic ratio, and
such a cancellation provides an interesting connection between
geometry,
equivalence principle and special renormalization properties
(cancellation of
strongest divergencies) for particles with $g=2$.
Another interesting connection is provided by the fact,
that it is just the deviation from $g=2$, which determine the
Gerasimov-Drell-Hearn sum rule for particle with arbitrary spin \cite{GDH}.

The crucial factor $1/2$ makes the evolution of the particle helicity
in magnetic and gravitomagnetic fields
rather different. The spin of the (Dirac) particle in the magnetic
field is dragging with the cyclotron frequency, being
twice larger than
Larmor one. It coincides with the frequency of the velocity
precession so that helicity is conserved. At the same time, the
gravitomagnetic field is making the velocity dragging twice
faster than spin, changing the helicity. This factor of $2$, however,
is precisely the one required by the possibility to reduce all the effect
of gravitomagnetic field to the frame rotation. While spin vector
is the same in the rotating frame and is
dragging only due to the rotation of the coordinate axis,
the velocity one is transformed and getting the
additional contribution, providing factor 2 to Coriolis acceleration.
The Dirac particle helicity conservation in magnetic field
allows to find semiclassical
interpretation of anomalous magnetic moment and axial anomaly \cite{Mas}.
The geometrical factor 1/2 is providing the contact of these
phenomena with the equivalence principle.
The similarity between gravitational and electromagnetic interactions,
leading to the simplifications of the field equations for $g=2$
was mentioned in \cite{Hripl}, although the low energy-theorem,
guaranteeing the appearance of
this value in the case of gravity, was not used.

Note that all the consideration is essentially based on the smallness
of the particle velocity, achieved by the mentioned Lorentz boost,
and therefore do not leading to the loss of generality.

Let us consider massive particle scattered by rotating astrophysical
object. The effect of the gravitomagnetic field is reduced
to the rotation of the local comoving frame, which is becoming
inertial at large distances before and after scattering.
Consequently, the helicity is not changed by gravitomagnetic field,
which is confirmed by the explicit calculation of the Born
helicity-flip matrix element in the case of massive neutrino \cite{HD2}.
The reason for that may be also deduced from the structure of the
matrix element (\ref{genc}), where, in general,
the symmetric combination of spins $(s+s')/2$ should appear, which
is zero in the spin-flip case.

Note that massless particles are not coupled to gravitomagnetic
field at all, as one cannot construct the pseudovector
in (\ref{genc}), because vectors $s$ and $p$ are collinear (c.f. \cite{HD}).
At the same time, for arbitrary light longitudinally polarized
massive particle, one
get the mass-independent term $P^{(\mu} \epsilon_{\perp}^{\nu) \alpha}
\Delta_{\alpha}$, where $\epsilon_{\perp}$ is a two-dimensional
antisymmetric tensor in the plane, orthogonal to particle momentum.

It may seem, that the equivalence principle should exclude
the possibility of helicity flip in the scattering by gravity source
at all. This is, however, not the case, if usual Newtonian-type
"gravitoelectric"
force is considered. Its action is also reduced to the local acceleration
of the comoving frame, in which the helicity of the particle is not altered.
However, the comoving frame after scattering differs from the initial one
by the respective velocity $\delta \vec v = \int \vec a dt$. The
corresponding boost to the original frame
is, generally speaking, changing the helicity
of the massive particle (the similar effect for the gravitomagnetic field
is just the rotation for the solid angle $\delta \vec \Omega =
\int \vec \omega dt$ and does not affect the helicity).
The same boost may be considered as a source of the famous deflection
of particle momentum $\delta \phi \approx |\delta \vec v|/|\vec v|$.
The average helicity of the completely polarized beam after such a
scattering may be estimated in the semiclassical approximation as
$<P> \approx cos \phi \approx 1-\phi^2/2$.
Due to the correspondence principle, this quantity may be expressed as
\begin{eqnarray}
<P>=\frac{d\sigma_{++}-d\sigma_{+-}}{d\sigma_{++}+d\sigma_{+-}} \approx
1-2 \frac{d\sigma_{+-}}{d\sigma_{++}},
\label{P}
\end{eqnarray}
where $d\sigma_{+-} \ll d\sigma_{++}$ - the helicity-flip and
non-flip cross-sections, respectively. Comparing "classical" and
"quantum" expression for $<P>$, one get
\begin{eqnarray}
\frac{d\sigma_{+-}}{d\sigma_{++}} \approx \frac{\phi^2}{4}
\label{+-}
\end{eqnarray}
To check this simple approach, one may perform the calculation
of this ratio for the Dirac particle scattered by the gravitational source.
In the Born approximation, the result is easy to find:

\begin{eqnarray}
\frac{d\sigma_{+-}}{d\sigma_{++}} \approx \frac{\phi^2}
{4(2\gamma-\gamma^{-1})^2}.
\label{+-e}
\end{eqnarray}

This expression is coinciding with the estimate (\ref{+-}),
as soon as
the particle is slow
($\gamma=E/m \to 1$), while for the fast particles
\begin{eqnarray}
\frac{d\sigma_{+-}}{d\sigma_{++}} \approx \frac{\phi^2}{16\gamma^2}.
\label{+-r}
\end{eqnarray}

Such an effect should, in particular, lead to the helicity flip
of any massive neutrino. It is very small, when
the scattering by the single object is considered,
but may be enhances while neutrino is propagating in Universe.
Should the propagation time be large enough, the effect would
result in unpolarized beam of the initially polarized neutrino,
effectively reducing its intensity by the factor of 2.

The manifestation of post-Newtonian equivalence principle
is especially interesting, when "gravitoelectric" component
is absent. Contrary to electromagnetic case, one cannot
realize this situation through cancellation of contributions of
positive and negative charges. At the same time, one may consider
instead the interior of the rotating shell (Lense-Thirring effect).
Especially interesting is the case of the shell constituting
the model of Universe, whose mass and radius are
of the same order, when the dragging frequency may be equal
to the shell rotation
frequency, which is just the Mach's principle \cite{MTW}.
One should note, that the low energy theorem, guaranteeing the
unique precession frequency for all quantum and classical rotators,
is the necessary counterpart of the Mach's principle.

\vskip\baselineskip
\noindent
{\bf 3 Universal nullification of anomalous gravitomagnetic
moment as an extension of equivalence principle.}

Up to this moment, we considered the gravitational interaction
of the particle, being the eigenstate
of the momentum and spin projection
and described by the
conserved energy-momentum tensor.
Any assumptions
on the particle locality except the locality of energy-momentum
tensor were unnecessary.

We are now ready to postulate the following straightforward generalization
of this principle:

{\bf Contributions of all fundamental constituents to the
Anomalous Gravitomagnetic Moment of composite particle are zero}

The main reason is the stability of the particle with respect
to action of gravitomagnetic field. To illustrate the latter, one may
consider the following gedankenexperiment. Suppose the (attractive)
interaction between two particles is adiabatically increasing,
so that they finally form the bound state. Originally the AGM of
both particles are equal to zero. When interaction is increasing,
the momenta and angular momenta of particles are no more conserved,
and this may generate, in principle, their non-zero AGM (equal to each other,
up to a sign). The gravitomagnetic field should force the particles spins
to rotate with the different frequencies,
affecting in that way the structure of the bound state.
If the bound state is "local" in the sense that its intrinsic
structure is never affected by gravity, such an opportunity
is excluded, and AGM is zero for each of interacting particles
separately.

The extension of this property to quarks and gluons, which
do not exist as a free particles, is equivalent to the
statement, that nucleon properties are not affected
by the gravity. More formally, this should mean that for each term
in the QCD action in the gravitational field, which may be transformed
to its flat-space form by the local coordinate transformation,
this local form would be sufficient to calculate the
relevant matrix element of the energy-momentum tensors.

The main consequence of such a hypothesis is a relation of two
rather different fields of physics: gravity and strong interactions.
The situation is somehow similar to the one existed
in hadron electrodynamics
before QCD. It was possible to establish some properties of
electromagnetic interactions of hadrons (and, consequently,
of matrix elements governed by strong interactions) starting
from gauge symmetry and corresponding Ward identities.

The existence of such a connection does not require a deep relation
like unified theory. At the same time, due to the generality of gravity
the correspondent "gravidynamics of hadrons" may provide more
elaborate links.

In any case, one may get some information about the
gravitational interactions of nucleons
performing the experimental and theoretical studies
of strong interactions.  Let us briefly outline the
main possible connections and directions.

The immediate problem one meet decomposing the quark and gluon
contributions to energy-momentum tensors is the appearance of the
structures (at the zero order in $\Delta$)
in (\ref{def})
proportional to $g^{\mu \nu}$.
The natural way of handling is the extraction
of the traceless part \cite{ji95}:

\begin{eqnarray}
      \langle p| T_{q,g}^{\mu\nu} |p\rangle
       &=&
      \langle p| T_{q,g}^{\mu\nu}-\frac{1}{4}T_{q,g}^{\mu\mu} |p\rangle
      +\langle p| \frac{1}{4}T_{q,g}^{\mu\mu} |p\rangle
\label{tr}
\end{eqnarray}

The traceless and trace parts are providing $3/4$ and $1/4$
of expectation value of $T^{00}$ component, related to the particle
{\it inertial} mass \cite{ji95}. At the same time, the interaction with the
external gravitational field (\ref{h}) is providing, due to its space
components $h_i^j=-2\phi\delta_i^j$, the respective contributions
$3/2$ and $-1/2$ to its {\it gravitational} mass. This sign difference
should not come as a surprise, because the Einstein equations differ just
by the sign, when the traces and traceless parts of the tensors
$T^{\mu \nu}$ and $R^{\mu \nu}$ are considered.

We are now in a position to postulate that it is just
{\it traceless} part of the forward matrix element,
which should be equal, due to equivalence principle,
to the linear in $\Delta$ part of the non-forward one.

To justify this choice, one might recall that it is the traceless
part allowing for the natural separation of the quark and gluon
contribution to the nucleon \cite{ji95}
mass. Also, the linear in $\Delta$ term (\ref{genc})
is manifestly traceless.
Another support is coming from the perturbative calculations.
The matrix elements of energy momentum tensors of
electrons and photons acquire the logarithmycally divergent contributions,
cancelled in their sum. This problem, at leading order(LO), is similar
to the calculation of QED corrections to gravity coupling \cite{Rob}.
It is sufficient to consider the matrix elements of either electron
or photon energy momentum tensor switched between free electron states,
and the latter case is more simple, being described by the single diagram.
It is enough to consider the terms of zero and first order in $\Delta$.
The divergent contribution to the former is appearing
in the traceless part and may be identified
with the second moment of spin-independent
DGLAP kernel $\int_0^1 dx x P_{Gq}(x)$. The linear term
is known from the orbital angular momentum calculations \cite{Ji,HS}
and is also equal to that quantity, so that AGM is really zero.


To summarize, the extension of the equivalence principle
for the fundamental fields in the constituent particle is:

\begin{eqnarray}
\langle P'| T_i^{\mu\nu} |P\rangle=N_i
[2 (P^{\mu} P^{\nu}-g^{\mu \nu}/4M^2) +
\frac{i}{M}P^{(\mu}\epsilon^{ \nu)  \sigma \rho \alpha}
P^\rho S^\sigma \Delta^\alpha]+O(g^{\mu \nu}, \Delta^2).
\label{genc}
\end{eqnarray}

As soon as the coefficient of the traceless part is equal to that
of $P^\mu P^\nu$, the trace term in square bracket may be omitted;
in particular, for quarks and gluons in nucleon  one get:

\begin{eqnarray}
B_{q,g}(0)=0; \\
P_{q, g} = J_{q, g}
\end{eqnarray}

\vskip\baselineskip
\noindent
{\bf 4 Discussion and Conclusions}

The main result of this paper is the relation between
gravity and strong interactions, analogous to the relation between
electrodynamics and strong interactions, provided by the
methods of electrodynamics of hadrons.
One should clearly distinguish the consideration of the stable composite
particle and its constituents. While in the first case
the zero AGM is a consequence
of the conservation of momentum and angular momentum,
the similar statement for its constituents should be considered
as an extension of equivalence principle. Let us briefly discuss the main
points for these related problems.

For composite particle as a whole,
the presented derivation of zero AGM allows to
understand the reason for the equal dragging frequencies for
macroscopic and microscopic rotators. This is due to the intimate relation
between the universal Dirac value of gyrogravitomagnetic ratio
(related, in turn, to the special renormalization properties
of the higher spin particles with $g=2$)
and geometrical factor for the effective gyrogravitomagnetic field.
One should not apply any notion of locality for the
particle in question, except that it should be the eigenstate
of the momentum and orbital angular momentum and, due to the
uncertainty principle, is {\it not} local in the real space;
one may recall the Dirac electron, which is point-like in
the dynamical sense, but may be described by the plane wave.
Also, the locality of energy momentum tensor is important.
The universality of gyrogravitomagnetic ratio is resolving the
ambiguity existing in the literature concerning the particles
motion in the external gravitomagnetic fields,
completing the proof of the gravitational Larmor theorem,
and constitute the necessary ingredient of Mach's principle.
In particular, the equivalence principle is not allowing for
helicity flip of any massive particle (say, neutrino)
due to the rotation of
massive astrophysical object passed by this particle.
At the same time, the Newtonian
"gravitoelectric" field is resulting in the helicity flip,
which may be related, in the semiclassical approximation,
to its deflection angle.

Passing to the structure of the composite particle,
one should find the nullification of its total AGM as a
new general constraint which should be imposed for {\it any} bound state,
and in particular for heavy or light nuclei. The simplest case
is provided by the deuteron, for which the isotopic structure is making
the AGM to be related to the usual AMM: namely, neglecting the sea quarks
contributions they are represented by the consecutive moments of
skewed parton distribution \cite{Ji,Rad}. Taking into account the
approximate cancellation of proton and neutron AMM (which may be explained
\cite{OT98} as resulting from the zero AGM of quarks in protons,
to be discussed below), one should find small AMM (to be manifested
in the experiments with polarized deuterons \cite{deut}) and
zero AGM. At the same time, for non-isosinglet nuclei,
there is no reason to expect the small AMM, while zero AGM
is providing the non-trivial constraint for their structure.
Moreover, the same reasoning may be applied for the atoms, as
soon as they are representing the {\it pure} quantum states,
and even to the coherent macroscopical structures in the
condensed matter physics.

The zero AGM of constituents and, moreover, of the fundamental
fields, is the new hypothesis, based on the locality of the
particle.
The specific role of traceless part of the energy-momentum tensor
might be compared to the modifications of gravity theory \cite{wein89}
providing the natural solution of cosmological constant problem.
The theoretical and experimental studies of the hadronic matrix
elements offer the possibility to understand the hadron
behavior in gravitational fields. From the other side,
the gravitational experiments \cite{expg} might be used
to understand the hadron structure.

\noindent
{\bf Acknowledgments.} I wish to thank A.V. Efremov, S.B. Gerasimov,
E. Leader, L. Mankiewicz, L. Masperi,
B. Pire, R. Ruskov, P. Ratcliffe and  A. Sch\"afer
for stimulating
discussions.

\end{document}